\documentclass[a4paper,12pt]{article}
\pdfoutput=1 

\newcommand{\nc}{\newcommand*} 

\nc{\al}{\alpha}
\nc{\s}{\sigma}
\nc{\dt}{\delta}
\nc{\Dt}{\Delta}
\nc{\Ld}{\Lambda}
\nc{\p}{\partial}
\nc{\Om}{\Omega}
\nc{\rd}{\mathrm{d}}
\nc{\Od}{\mathcal{O}} 

\def\({\left(}
\def\){\right)}
\def\[{\left[}
\def\]{\right]}
\def\e{\begin{equation}}
\def\q{\end{equation}}
\def\m{\begin{eqnarray}}
\def\n{\end{eqnarray}}

\nc{\Eq}[1]{Eq.~\eqref{#1}}     
\nc{\Fig}[1]{Fig.~\ref{#1}}     
\nc{\Table}[1]{Table~\ref{#1}}  
\nc{\Sec}[1]{Sec.~\ref{#1}}     

\nc{\Msun}{M_\odot}             
\nc{\fpbh}{f_{\mathrm{pbh}}}    
\nc{\fpbhn}{f_{\mathrm{PBH0}}}    
\nc{\mR}{\mathcal{R}} 
\nc{\seq}{\sigma_{\mathrm{eq}}}
\nc{\ogw}{\Omega_{\mathrm{GW}}}
\nc{\gpcyr}{\mathrm{Gpc}^{-3}\,\mathrm{yr}^{-1}}
\nc{\lvc}{LIGO/Virgo} 
\nc{\SNR}{\mathrm{SNR}} 
\nc{\mmin}{{m_{\mathrm{min}}}}
\nc{\mmax}{{m_{\mathrm{max}}}}
\nc{\Mmin}{{M_{\mathrm{min}}}}
\nc{\fmin}{{f_{\mathrm{min}}}}
\nc{\VT}{\mathrm{VT}}
\nc{\rhoGW}{\rho_{\mathrm{GW}}}
\nc{\vth}{\vec{\theta}}
\nc{\vd}{\vec{d}}
\nc{\vla}{\vec{\lambda}}
\nc{\Nobs}{N_{\mathrm{obs}}}
\nc{\av}[1]{\langle #1 \rangle} 
\nc{\km}{\mathrm{km}}
\nc{\Mpc}{\mathrm{Mpc}}
\nc{\Tobs}{T_{\mathrm{obs}}}
\nc{\Ntemp}{N_{\mathrm{temp}}}
\nc{\cH}{{\mathcal{H}}}

\nc{\OL}{{\Omega_\Lambda}}
\nc{\OM}{{\Omega_\mathrm{m}}}
\nc{\OR}{{\Omega_r}}
\nc{\mU}{{\mathcal{U}}}
\nc{\Mc}{{M_\mathrm{c}}}
\nc{\sgc}{{\sigma_\mathrm{c}}}
\nc{\Mf}{{M_\mathrm{f}}}
\nc{\LCDM}{{$\Lambda$CDM}}
\nc{\kmsmpc}{km~s$^{-1}$~Mpc$^{-1}$}

\nc{\addref}{[\textcolor{red}{add ref}] } 
\nc{\eg}{\textit{e.g.~}}
\nc{\app}{\approx}
\nc{\hf}{\frac{1}{2}}
\nc{\discuss}{\textcolor{red}{Add discussion here!}}
\nc{\red}[1]{\textcolor{red}{#1}}
\usepackage[symbol]{footmisc}
\usepackage{jcappub} 
\usepackage{changes}
\usepackage{amsmath}
\usepackage{graphicx}
\usepackage{adjustbox}
\usepackage{multirow}
\usepackage[normalem]{ulem}

\title{Constraints on peculiar velocity distribution of binary black holes using gravitational waves with GWTC-3}

\author{Zhi-Qiang You,$^{1,2,3}$}
\author{Zu-Cheng~Chen,\note{Corresponding author.}$^{4,5,1,2,*}$}
\author{Lang~Liu,$^{1,2,*}$}
\author{Zhu~Yi,$^{2}$}
\author{Xiao-Jin~Liu,$^{1,2}$}
\author{You~Wu,$^{6}$}
\author{and Yi~Gong$^{7}$}

\affiliation{$^1$Department of Astronomy, Beijing Normal University, Beijing 100875, China}
\affiliation{$^2$Advanced Institute of Natural Sciences, Beijing Normal University, Zhuhai 519087, China}
\affiliation{$^3$Henan Academy of Sciences, Zhengzhou 450046, Henan, China}
\affiliation{$^4$Department of Physics and Synergetic Innovation Center for Quantum Effects and Applications, Hunan Normal University, Changsha, Hunan 410081, China}
\affiliation{$^5$Institute of Interdisciplinary Studies, Hunan Normal University, Changsha, Hunan 410081, China}
\affiliation{$^6$College of Mathematics and Physics, Hunan University of Arts and Science, Changde, 415000, China}
\affiliation{$^7$School of Physics and Technology, Wuhan University, Wuhan, Hubei 430072, China}

\emailAdd{you\_zhiqiang@whu.edu.cn}
\emailAdd{zuchengchen@hunnu.edu.cn}
\emailAdd{liulang@bnu.edu.cn}
\emailAdd{yz@bnu.edu.cn}
\emailAdd{xliu.astro@bnu.edu.cn}
\emailAdd{youwuphy@gmail.com}
\emailAdd{gongyi@whu.edu.cn}
\abstract{
Peculiar velocity encodes rich information about the formation, dynamics, evolution, and merging history of binary black holes. In this work, we employ a hierarchical Bayesian model to infer the peculiar velocity distribution of binary black holes. We use the data from GWTC-3 and assume a Maxwell-Boltzmann distribution for the peculiar velocities, but do not consider the dependence of peculiar velocity on the masses of black hole binaries. The constraint on the peculiar velocity distribution parameter, $v_0$, is weak and uninformative. However, the determination of peculiar velocity distribution can be significantly improved with next-generation ground-based gravitational wave detectors. For the Einstein Telescope, the relative uncertainty of $v_0$ will reduce to $\sim$ 10\% using $10^3$ golden binary black hole events. Our statistical approach thus provides a robust and prospective inference for determining the peculiar velocity distribution.
}

\begin{document}
\maketitle
\flushbottom

\section{\label{intro}Introduction}

Peculiar velocities are so sensitive to the matter distribution on large scales that we can use them as a useful cosmological probe and to test the link between gravity and matter~\cite{vp_sn_2014MNRAS.442.1117D}.
The peculiar velocities observed in gravitational wave (GW) events may result from various astrophysical effects, such as {binary motion in a Galactic potential}, kicks imparted by supernovae during binary evolution, the peculiar motion of the host galaxy relative to the Hubble flow, and the velocity predicted by the linear momentum dissipated through GW emission according to general relativity. Importantly, during the early phase of the inspiral at a reference frequency of 20~Hz, the velocity caused by linear momentum dissipated by GW emission is relatively insignificant compared to other astrophysical factors~\cite{Blanchet:2005rj,LIGOScientific:2017adf, Hjorth:2017yza}.
The peculiar velocity, predicted by the linear momentum emission, can be inferred using the recoil velocity of remnant black holes~\cite{2007PhRvL..98i1101G,2011PhRvL.107w1102L} and a simple method given by Ref.~\cite{2018PhRvL.121s1102C}.
Based on this method~\cite{2016PhRvL.117a1101G,2018PhRvL.121s1102C,2020PhRvL.124j1104V}, the recoil velocity of the binary black hole (BBH) merger GW200129~\cite{LIGOScientific:2021djp} can reach $\sim1542$ $\mathrm{km}\,\mathrm{s^{-1}}$~\cite{2022PhRvL.128s1102V}.
Actually, the peculiar velocity evolves across the inspiral, merger, and ringdown phases and can, in principle, be inferred by observing the differential Doppler shift throughout the GW signal~\cite{2016PhRvL.117a1101G}.
In this work, we study the average peculiar velocity of GW binaries at a reference frequency $f_\mathrm{ref}=20\, \mathrm{Hz}$ across the population.

Recently, as advanced LIGO~\cite{aligo_2015CQGra..32g4001L} and Virgo~\cite{avirgo_2015CQGra..32b4001A} improve their sensitivities and the fourth observing run starts, unprecedented opportunities arise to study the astrophysics of black holes, including directly measuring the peculiar velocities of GW sources. However, direct measurement of peculiar velocity depends sensitively on the signal-to-noise ratio (SNR) of an event, as recoil velocity can only be measured when the SNR is sufficiently high and by using a waveform calculated by numerical relativity surrogate models~\cite{2019PhRvR...1c3015V,2019PhRvL.122a1101V,2019PhRvD..99f4045V,2017PhRvD..96b4058B,NR_waveform_Varma:2020bon,spin_Biscoveanu:2021nvg,spin_Varma:2021csh,super_kick_Ma:2021znq}.
Unfortunately, BBH events with large SNR are rare, and in GWTC-3 only three BBH events have an SNR $> 20$, including GW200129~\cite{LIGOScientific:2021djp}.
Estimating the peculiar velocity of most BBH events, which usually have low SNRs, remains an interesting problem.

As demonstrated in Ref.~\cite{LIGOScientific:2017adf}, peculiar velocity is an important input parameter in constraining the Hubble constant, $H_0$. Hence, one can constrain the peculiar velocity for a given cosmological model, providing an alternative method for inferring the average peculiar velocity of BBHs. For a flat $\Lambda$CDM model with fixed cosmological parameters, the peculiar velocity encodes in the correlation between the redshifted mass and the luminosity distance distributions of black holes, which have no electromagnetic counterparts or host galaxy.
Indeed, Ref.~\cite{vp_sn_2014MNRAS.442.1117D} has shown the relationship between the peculiar velocity and the resultant Doppler shift.
For the BBH events detected by current GW observatories, due to the significant uncertainties in the parameter estimations, the results of constraints on peculiar velocity distribution may not be ideal.
However, in the era of third-generation ground-based detectors, such as the Einstein Telescope (ET) \citep{2010CQGra..27s4002P} and Cosmic Explorer (CE) \citep{abbott2017exploring}, a significant improvement in BBH sample size and parameter estimation accuracy is expected.
It will help to enhance the result of peculiar velocity parameter inference because the ET/CE detector will be able to detect BBH mergers throughout the Universe, yielding $\sim 10^4 - 10^7$ discoveries per year \citep{2011arXiv1108.1423S,Chen:2019irf}. 

In this work, we simultaneously infer the peculiar velocity distribution and BBH population properties utilizing GWTC-3 and mock data from ET.
This paper is organized as follows. In Section~\ref{sec:mass}, we describe our model of BBH mass distribution. In Section~\ref{sec:cosmoloy_vp}, we introduce the redshift caused by the peculiar velocity of BBHs. Section~\ref{sec:bayes} reviews the Bayesian hierarchical inference method used for analysis, while Section~\ref{sec:results} shows the constraints on the peculiar velocity distribution. Finally, we summarize and discuss our results in Section~\ref{sec:discussion}. 
In addition, we show full posteriors using BBH events from GWTC-3 in Appendix~\ref{append}.
Throughout this work, we have set the speed of light to unity, namely $c = 1$.


\section{\label{sec:mass}BBH mass distribution}	

In this work, we introduce a method for jointly constraining peculiar velocity and source population properties of BBHs, without resorting to host galaxy information~\cite{2021PhRvD.104f2009M}. 
We use a phenomenological mass model, {\tt{POWER LAW + PEAK}}, in the source frame for BBHs~\cite{2021ApJ...913L...7A,LIGOScientific:2021aug}, that is preferred by GWTC-3~\cite{LIGOScientific:2021aug}. 
The probability distribution of primary black hole mass 
is $\pi(m_1) = \pi(m_1 | \alpha, m_{\rm{min}} , m_{\rm{max}}, \lambda_g, \mu_g, \sigma_g, \delta_m)$, namely
 \begin{equation}
\pi(m_1)=\left[(1-\lambda_g) \mathcal{P}(m_1 | m_{\min }, m_{\max },-\alpha)+\lambda_g \mathcal{G}(m_1 | \mu_{{g}}, \sigma_{{g}})\right] S(m_1 | m_{\min }, \delta_m),
\end{equation}
where the truncated power law $\mathcal{P}(m | m_{\min }, m_{\max }, \alpha)$ with spectral index $\alpha$, lower cut-off $m_{\rm{min}}$ and upper cut-off $m_{\rm{max}}$ is described as,
\begin{equation}
\mathcal{P}(m | m_{\min }, m_{\max }, \alpha) = \begin{cases}m^\alpha & (m_{\min } \leqslant m \leqslant m_{\max }), \\ 0 & \text{Otherwise}.\end{cases}
\end{equation}
Meanwhile, the Gaussian component with mean $\mu_g$ and standard deviation $\sigma_g$ is given by
\begin{equation}
\mathcal{G}(m | \mu_g, \sigma_g)=\frac{1}{\sigma_g \sqrt{2 \pi}} \exp \left[-\frac{(m-\mu_g)^2}{2 \sigma_g^2}\right].
\end{equation}
The ratio of truncated power law and Gaussian component is $\lambda_g$. And the smoothing function, which rises from 0 to 1 over interval $(m_{\min }, m_{\min }+\delta_m)$~\cite{2021ApJ...913L...7A}, is written as
\begin{equation}
S(m | m_{\min }, \delta_m)= \begin{cases}
0 & (m<m_{\min }), \\ 
{\left[f(m-m_{\min }, \delta_m)+1\right]^{-1}} & (m_{\min } \leq m<m_{\min }+\delta_m), \\ 
1 & (m \geq m_{\min }+\delta_m),
\end{cases}
\end{equation}
with 
\begin{equation}
f(m^{\prime}, \delta_m)=\exp(\frac{\delta_m}{m^{\prime}}+\frac{\delta_m}{m^{\prime}-\delta_m}).
\end{equation}
The secondary black hole mass distribution $\pi (m_2)=\pi (m_2 | m_1, m_{\rm{\min}},\beta, \delta_m)$,  given the primary black hole mass $m_1$, is constructed by a truncated power-law with index $\beta$ between a minimum mass $m_{\rm{min}}$ and a maximum mass $m_1$, and a smoothing  function $S(m_2 | \delta_m, m_{\min })$,
\begin{equation}
\pi(m_2)=\mathcal{P}(m_2 | m_{\min }, m_1, \beta) S(m_2 | m_{\min }, \delta_m).
\end{equation}
Based on the above discussion, the source-frame mass distribution for the BBH population is
\begin{equation}
\pi(m_1, m_2 | \Phi_m)=\pi(m_1 | \alpha, m_{\min }, m_{\max }, \lambda_g, \mu_g, \sigma_g, \delta_m)\, \pi( m_2 | m_1, m_{\rm{\min}},\beta, \delta_m ),
\end{equation}
where $\Phi_m = \{\alpha, \beta, m_{\min }, m_{\max }, \lambda_g, \mu_g, \sigma_g, \delta_m \}$ are the parameters that describe the BBHs mass distribution for {\tt{POWER LAW + PEAK}} model.

\section{Redshift induced by peculiar velocity\label{sec:cosmoloy_vp}}
The mass and luminosity distance of GW events contain redshift information. The GW observatories detect redshifted black hole masses $m^{\rm{det}}_i$ and luminosity distance $D_L$ correlated with source-frame masses $m_i$ and redshift by
\begin{equation}\label{eq:ms-mdet}
m_i=\frac{m_i^{\text{det}}}{1+z} .
\end{equation}
We infer the redshift of GW events and use the spectral siren method by tracking the mass spectrum in different luminosity distance bins for studying cosmic expansion and peculiar velocity of black holes.

In this work, we assume a flat Friedmann-Robertson-Walker Universe and a $\Lambda$CDM model by fixing cosmological parameters since we are interested in peculiar velocities. The Hubble rate at redshift $z$ is 
\begin{equation}
H(z)=H_0 E(z),
\end{equation}
where $H_0$ is the Hubble constant, and 
\e 
E(z) = \sqrt{\OM (1+z)^3 + \OL}.
\q 
The $\Omega_m$ and $\Omega_{\Lambda}$ represent the matter density and the dark energy density, respectively.
In this work, we use the best-fit values of cosmological parameters from Planck 2018~\citep{Planck:2018vyg}, where $H_0=67.4 \, \mathrm{km} \,\mathrm{s}^{-1} \,\mathrm{Mpc}^{-1}$, $\OM=0.315$, and $\OL = 1 - \OM$. 
Given luminosity distance $D_L$ by GW observation, one can then calculate the redshift caused by cosmic expansion as
\begin{equation}\label{eq:dl-z}
D_L = \frac{(1+z)}{H_0} \int_{0}^{z} \frac{dz'}{E(z')},
\end{equation}
without considering the peculiar velocities of GW sources.

However, if the object has a peculiar velocity, an observer sees a 
Doppler effect in the cosmic microwave background, and the object acquires an additional peculiar redshift component $z_p$. Note that the peculiar velocity is the sum of the projections of the peculiar velocities of two objects in binary systems along the line of sight from the observer and sources, i.e., the radial velocity component. The redshift of the GW sources with non-zero peculiar velocities can be described as,
\begin{equation}\label{eq:zp_ztot}
1+z=(1+z_{cos})(1+z_p),
\end{equation}
where the $z_{cos}$ is the redshift caused by Hubble flow, i.e., recession velocity, which can be calculated by \Eq{eq:dl-z}, and $z_{p}$ is the redshift caused by peculiar velocities of GW sources. In Ref.~\cite{vp_sn_2014MNRAS.442.1117D}, the relationship between $z_p$ and peculiar velocity $v_p$ is 
\begin{equation}\label{eq:vp_zp}
1+z_p=\sqrt{\frac{1+v_p}{1-v_p}}.
\end{equation} 
When $v_p$ is much slower than the speed of light ($v_p \ll 1$), \Eq{eq:vp_zp} can be approximated by $1+z_p \approx 1+v_p$. 
In this work, we simply assume the peculiar velocities follow a Maxwell-Boltzmann distribution as~\cite{Bird:2016dcv}
\begin{equation}
\pi(v_p|v_0) \propto \frac{v_p^2 }{v_0^3}\exp (-\frac{v_p^2}{v_0^2} ),
\end{equation}
where $v_p$ is the peculiar velocity of each BBH event, and $v_0$ is the most probable velocity of all BBH events. We set $v_0$ as a free parameter.

\section{Hierarchical Bayesian inference\label{sec:bayes}}
In this section, we introduce  hierarchical Bayesian inference to estimate posterior distributions of parameters that describe the
 motion and the source population properties of BBHs. 
 We use a cosmic star formation rate (SFR) model from~\cite{madau2014cosmic} to describe the evolution of BBH binary merger rate since the black hole formation rate might track the SFR. The component $\psi(z | \gamma, k, z_{p})$ indicating the relation between BBH redshift and the merger rate is~\cite{madau2014cosmic,LIGOScientific:2021aug}
 \begin{equation}
 \psi(z | \gamma, k, z_{\mathrm{pk}})=\left[1+(1+z_{\mathrm{pk}})^{-\gamma-k}\right] \frac{(1+z)^\gamma}{1+\left[(1+z) /(1+z_{\mathrm{pk}})\right]^{\gamma+k}},
 \end{equation}
with a peak at $z_{\rm{pk}}$, where $\gamma$ and $\kappa$ are the power-law indexes at low redshift and high redshift interval, respectively. 
Then the merger rate in the detector frame can be written as~\cite{2020ApJ...896L..32C}, 
\begin{equation}\label{eq:merger_rate}
\mathcal{R}(z) = R_0 \pi(z| \gamma,\kappa,z_{\rm{pk}}),
\end{equation}
with
\begin{equation} \label{eq:redshift_prior}
 \pi(z| \gamma,\kappa,z_{\rm{pk}})=C \frac{1}{1+z} \frac{d V_{\mathrm{c}}}{d z} \psi(z | \gamma, k, z_{\mathrm{pk}}). 
\end{equation}
\Eq{eq:merger_rate} gives the merger rate as a function of redshift. $R_0$ is the local merger rate, which is the integral of \Eq{eq:merger_rate} over other observed parameters given at $z=0$. 
In the redshift prior function \Eq{eq:redshift_prior}, $C$ is the normalization factor, and $d V_{\mathrm{c}} / d z$ is the differential comoving volume. In order to convert the clock from the source frame to the detector frame, we multiply the factor ${1}/{(1+z)}$ by the redshift prior. 

We have introduced the source mass and peculiar velocity distribution in Section~\ref{sec:mass} and Section~\ref{sec:cosmoloy_vp}. The complete characterization of the population models is the product of redshift, mass, and peculiar velocity prior, written as 
\begin{equation}\label{eq:ppop}
p_{\text{pop}}(\theta | \Phi_m, \gamma ,\kappa,z_{\rm{pk}}, v_0 ) = C \frac{1}{1+z} \frac{d V_{\mathrm{c}}}{d z} \psi(z | \gamma, k, z_{\rm pk}) \pi (m_1,m_2 | \Phi_{m} )  \pi(v_p | v_0),
\end{equation}
where $\Phi_m = \{\alpha, \beta, m_{\min }, m_{\max }, \lambda_g, \mu_g, \sigma_g, \delta_m \}$   is the {\tt{POWER LAW + PEAK}} mass population parameters not related to cosmology and peculiar velocity, and $\theta = \{ m_1,m_2,z,v_p \}$ are the intrinsic GW parameters. All parameters introduced above are shown in Table~\ref{tab:priorss}.

We are concerned with the distribution of BBH source parameters $\theta$ conditional upon the hyperparameters $(\Phi_m, \gamma, \kappa,z_{\rm{pk}}, v_0 )$, i.e. conditional prior $p_{\text{pop }}(\theta | \Phi_m, \gamma,\kappa,z_{\rm{pk}}, v_0 )$, by analysing $N_{\mathrm{obs}}$ GW data $\textbf{h} = \{h_1, h_2, \cdots, h_{{N_{\rm{obs}}}}\}$. 
Accordingly, the hyper-likelihood $\mathcal{L}(\mathbf{h}  | \Phi_m, \gamma ,\kappa,z_{\rm{pk}}, v_0 , R_0)$ related to the single-event likelihood $\mathcal{L}(h_i | \theta)$ for the $i$-th event can be written as~\cite{Loredo:2004nn,Thrane:2018qnx,Mandel:2018mve,Chen:2018rzo,Chen:2021nxo,Liu:2022iuf,Zheng:2022wqo}
\begin{align}\label{eq:likelihood}
\mathcal{L}(\mathbf{h}  | \Phi_m, \gamma ,\kappa,z_{\rm{pk}}, v_0 , R_0   ) & \propto R_0^{N_{\text{obs}}} e^{-R_0 \xi(\Phi_m,\gamma ,\kappa,z_{\rm{pk}}, v_0) } \\ \nonumber
&  \times  \prod_{i=1}^{N_{\text{obs}}} \int \mathcal{L}(h_i | \theta) p_{\text{pop}}(\theta | \Phi_m, \gamma ,\kappa,z_{\rm{pk}}, v_0 ) d \theta ,
\end{align}
where $\xi(\Phi_m,\gamma, \kappa,z_{\rm{pk}}, v_0)$ is the detection fraction that quantifies selection biases for a population based on hyperparameters $(\Phi_m,\gamma, \kappa,z_{\rm{pk}}, v_0)$, described as
\begin{equation}\label{xi}
\xi(\Phi_m,\gamma ,\kappa,z_{\rm{pk}}, v_0)=\int P_{\mathrm{det}}(\theta)\, p_{\mathrm{pop}}(\theta | \Phi_m, \gamma,\kappa,z_{\rm{pk}}, v_0) d \theta .
\end{equation}
The detection probability based on the source parameters $\theta$ is $P_{\text{det}}(\theta)$, which is related to the simulated signals. 

In this work, the simulated injections in~\cite{ligo_scientific_collaboration_and_virgo_2021_5546676} are used to compute the detection fraction $\xi(\Phi_m,\gamma, \kappa,z_{\rm{pk}}, v_0)$.
We utilise a Monte Carlo integral over found injections~\cite{LIGOScientific:2021psn} to simplify \Eq{xi} as
\begin{equation}\label{eq:selection}
	\xi(\Phi_m,\gamma ,\kappa,z_{\rm{pk}}, v_0) \approx \frac{1}{N_{\mathrm{inj}}} \sum_{j=1}^{N_{\text{found }}} \frac{p_{\mathrm{pop}}(\theta_{j} | \Phi_m,\gamma ,\kappa,z_{\rm{pk}}, v_0)}{p_{\mathrm{draw}}(\theta_j)}.
\end{equation}
In practice, we calculate the sum of the conditional prior divided by $p_{\mathrm{draw}}$ for each event instead of the integral. 
$p_{\text{draw }}$ is the regular prior distribution from which the injections are drawn, where $j$ indicates the $j$-th detected event. 
We employ the package, \texttt{ICAROGW}~\cite{Mastrogiovanni:2021wsd} including the BBH population distribution \Eq{eq:ppop}, to estimate the likelihood function \Eq{eq:likelihood}, and then utilize \texttt{dynesty}~\cite{Speagle:2019ivv} sampler in \texttt{Bilby} package~\cite{Ashton:2018jfp,Romero-Shaw:2020owr} to estimate the hyperparameters.

\begin{table*}[htbp!]
	\centering
	\setlength\tabcolsep{0.9pt}
	\begin{adjustbox}{width=1.\textwidth}
		\begin{tabular}{llr}
			\hline\hline
			\textbf{Parameter} & \multicolumn{1}{c}{\textbf{Description}} & \textbf{Prior} \\
			\hline
                \multicolumn{3}{c}{\textbf{Peculiar velocity distribution}} \\[1pt]
                $v_0$  & The most probable peculiar velocity. & ${\rm{log}}\mU(-5, -2)$ \\ \hline
                \multicolumn{3}{c}{\textbf{Merger rate evolution}} \\[1pt]
			$R_0$ & local merger rate of BBHs in $\mathrm{Gpc}^{-3} \mathrm{yr}^{-1}$. & $\mU(0, 200)$\\         
			$\gamma$ & Power-law index for the rate evolution before the point $z_{\mathrm{pk}}$. & $\mU(0, 12)$\\
			$\kappa$ & Power-law index for the rate evolution after the point $z_{\mathrm{pk}}$. & $\mU(0, 6)$\\
			$z_{\mathrm{pk}}$ & Redshift turning point between the powerlaw regimes with $\gamma$ and $\kappa$. & $\mU(0, 4)$\\ \hline
                \multicolumn{3}{c}{{\tt{POWER LAW + PEAK}} \textbf{mass function}}\\[1pt]
			$\alpha$ & Spectral index for the power-law of the primary mass distribution. & $\mU(1.5, 12)$\\
			$\beta$ & Spectral index for the power-law of the mass ratio distribution. & $\mU(-4, 12)$\\
			$m_{\min}\,[\Msun]$ & Minimum mass of the primary mass distribution. & $\mU(2, 10)$\\
			$m_{\max}\,[\Msun]$ & Maximum mass of the primary mass distribution. & $\mU(50, 200)$\\
			$\lambda_g$ & Fraction of the model in the Gaussian component. & $\mU(0, 1)$\\
			$\mu_g\,[\Msun]$ & Mean of the Gaussian component in the primary mass distribution. & $\mU(20, 50)$\\
			$\sigma_g\,[\Msun]$ & Width of the Gaussian component in the primary mass distribution. & $\mU(0.4, 10)$\\
			$\delta_m\,[\Msun]$ & Range of mass tapering at the lower end of the mass distribution. & $\mU(0, 10)$\\ 
			\hline
		\end{tabular}
	\end{adjustbox}
	\caption{\label{tab:priorss} Parameters and their prior distributions used in the Bayesian parameter estimations. Note that $z_p$ and $z_{\rm{pk}}$ indicate the redshift caused by peculiar velocity and the redshift turning point of the merger rate evolution, respectively.}
\end{table*}

\section{\label{sec:results}Results}

In this section, we show the results of hyperparameter estimation using the BBHs from GWTC-3, and the simulated BBH population expected to be detected by third-generation detectors such as ET.

\subsection{GWTC-3}

Similar to Refs.~\cite{LIGOScientific:2021aug,Chen:2022fda}, we use 42 BBH events with $\mathrm{SNR}>11$ and inverse false alarm rate higher than four years, excluding two BNS events~\cite{gw170817_2017PhRvL.119p1101A,gw190425_2020ApJ...892L...3A}, two NSBH events~\cite{gw200105_nsbh_2021ApJ...915L...5A} and one asymmetric mass binary GW190814~\cite{gw190814_2020ApJ...896L..44A}, from GWTC–3. We use combined posterior samples estimated by the IMRPhenom~\cite{Thompson:2020nei,Pratten:2020ceb} and SEOBNR~\cite{Ossokine:2020kjp,Matas:2020wab} waveform families.
Based on the parameters estimation of these BBH sources, we perform data collection for the distribution of detector-frame masses and luminosity distance. 
Unlike the works interested in cosmological parameters estimation~\cite{farr2019future,LIGOScientific:2017adf,LIGOScientific:2021aug,spectrum_siren_2022PhRvL.129f1102E,You:2020wju,chen_2023JCAP...03..024C}, we are primarily concerned with peculiar velocity parameter of the BBH population.
Based on the relationship between peculiar velocity and redshift, we employ Eqs. \eqref{eq:dl-z}-\eqref{eq:vp_zp} to calculate peculiar velocity given the Hubble constant $H_0$ and matter density $\OM$~\cite{Planck:2018vyg}. We show all parameters and their prior interval utilized for Bayesian parameter estimations in Table~\ref{tab:priorss}. Generally, these parameters can be divided into four categories: peculiar velocity distribution, merger rate evolution, {\tt{POWER LAW + PEAK}} mass function, and cosmology.

We show extra details on the joint estimation of full posteriors in Figure~\ref{fig:all_para} in Appendix~\ref{append}, including the peculiar velocity distribution parameter, merger rate evolution parameters, and mass function parameters using BBHs.
For these corner plots of hyperparameters, most of the one-dimensional posterior for mass function and merger rate evolution parameters have the characteristics of Gaussian-like distribution except the peculiar velocity distribution parameter $v_0$.    
Figure~\ref{fig:gwtc_et_post_v0} shows the marginal posterior distributions for the peculiar velocity parameters $v_{0}$. 
For comparison, we pick up a portion of the one-dimensional marginal posterior distribution for $v_0$ in Figure~\ref{fig:all_para} to present as the blue histogram in Figure~\ref{fig:gwtc_et_post_v0}.  
There is no difference between the prior and posterior distributions, and they are flat distributions at the logarithmic scale. Therefore, the peculiar velocity distribution posterior calculated from GWTC-3 is uninformative, indicating the current BBH events detected by LIGO-Virgo-KAGRA can not constrain the peculiar velocity distribution.

\subsection{ET mock data}
The posteriors derived using GWTC-3 in this work, shown in Figure~\ref{fig:all_para}, are consistent with the result in Ref.~\cite{LIGOScientific:2021aug, 2023PhRvX..13a1048A}. 
To generate mock data for ET, we set the injected values based on our posterior results for {\tt{POWER LAW + PEAK}} mass model parameters $\alpha=3.8$, $\beta=0.8$, $m_{\max }=100 \mathrm{M}_{\odot}$, $m_{\min }=5 \mathrm{M}_{\odot}$, $\delta_m=4.8 \mathrm{M}_{\odot}$, $\mu_g=33 \mathrm{M}_{\odot}$, $\sigma_g=3.2 \mathrm{M}_{\odot}$, and $\lambda_g=0.03$, for the rate evolution model parameters $\gamma=4.7$, $k=3$, and $z_{\rm{pk}}=2.3$, and for BBH merger rate parameter
$R_0=17 \,\mathrm{Gpc}^{-3}\, \mathrm{yr}^{-1}$ at $z=0$. 
For the median value of peculiar velocity distribution $v_0$, we use $v_0=3.33\times 10^{-3}$ (corresponding to $1000\,\mathrm{km}\,\mathrm{s}^{-1}$) following Ref.~{\cite{Bird:2016dcv,LIGOScientific:2021djp}}. 
The third-generation detectors, such as ET, are expected to detect approximately $\sim$ $10^4$ golden events for BBHs per year with $\leq 10\%$ error on luminosity distance and $\leq 0.02\%$ error on masses~\cite{ET_since}.
We conservatively choose $10^3$ golden BBH events from these data sets, indicating $N_{obs} =10^3 $.
We adopt a Fisher matrix approach to evaluate the correlation between mass and distance while generating posterior distributions for the individual golden binary. We consider golden binaries with a relatively large SNR threshold of $100$ following~\cite{Branchesi:2023mws}.
To simulate Gaussian-type posteriors for individual events, we generate the peak value for each binary parameter posterior from a Gaussian distribution centered on the true value, with the uncertainty determined by the Fisher matrix. Subsequently, the standard deviations are calculated accordingly using the Fisher matrix method.

\begin{figure}[htbp!]
    \centering
    \begin{minipage}{0.45\textwidth}
        \raggedleft
        \includegraphics[width=\textwidth]{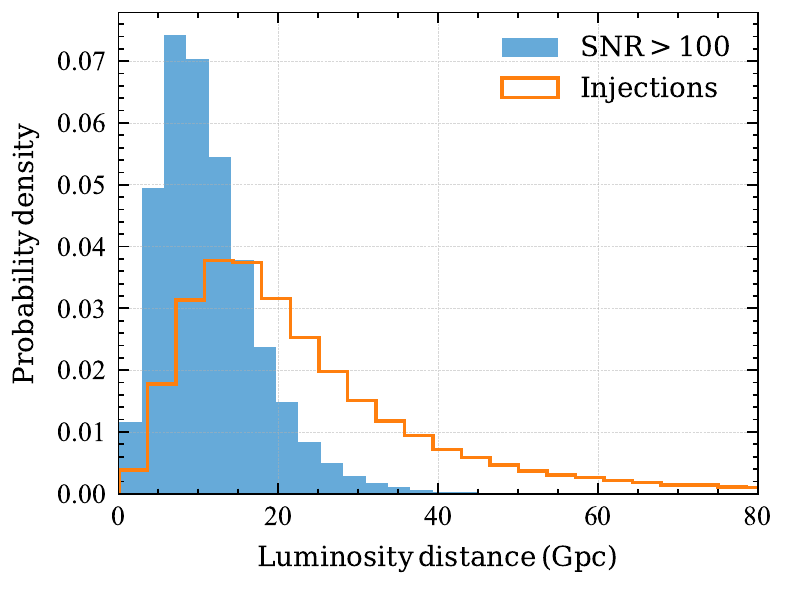}
        \includegraphics[width=\textwidth]{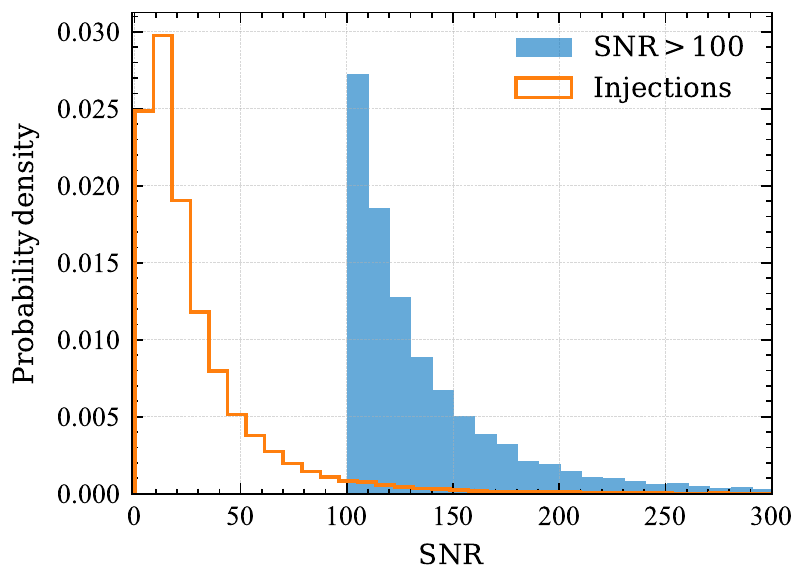}
    \end{minipage}
    \begin{minipage}{0.45\textwidth}
        \raggedright
        \includegraphics[width=0.958\textwidth]{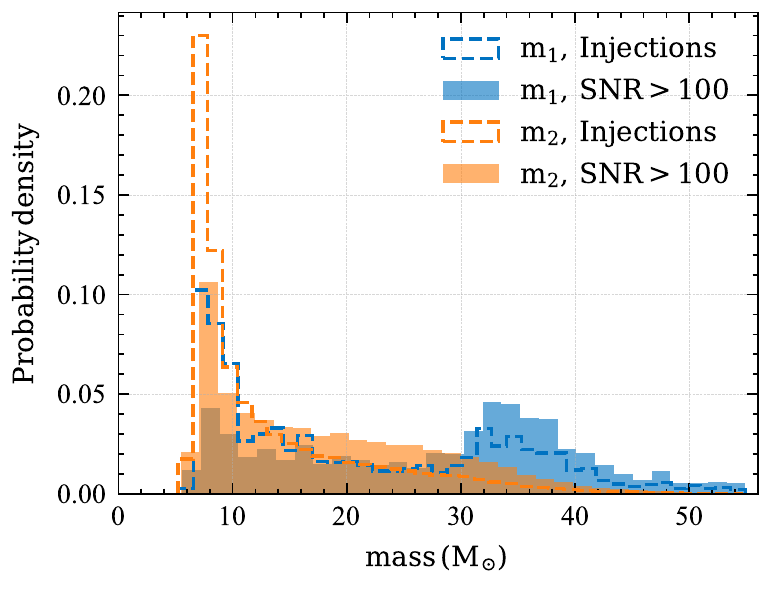}
        \includegraphics[width=0.977\textwidth]{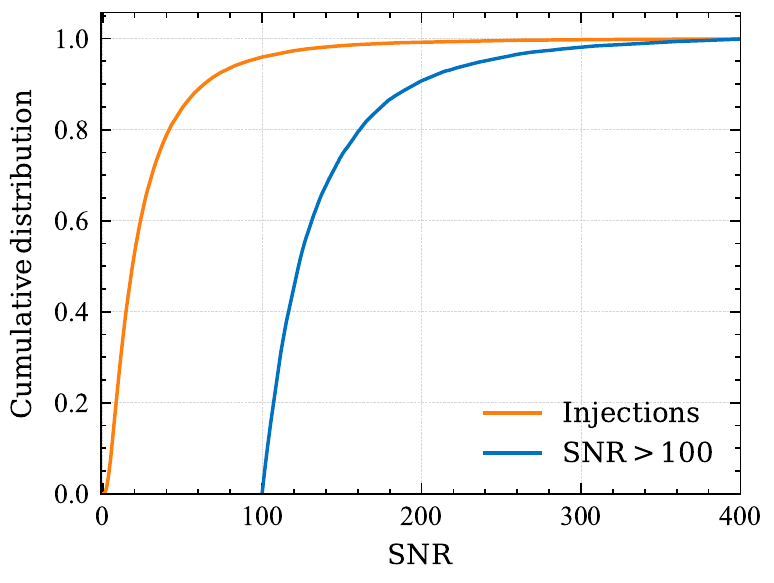}
    \end{minipage}
    \caption{\label{fig:et_inj} The distribution of luminosity distance, black hole mass ($m_1$, $m_2$), and SNR between the injection and the found events (SNR $>$ 100).}
\end{figure}

To evaluate the selection effect, { we simulate $N_\mathrm{inj}=10^7$ mock BBH events} and classify events with SNR $>100$ as the found events.
The luminosity distance, mass, and SNR distributions for both the injected BBH events and the detected events (SNR $>$ 100) are illustrated in Figure~\ref{fig:et_inj}. 
We employ the ET-D sensitivity curve~\cite{2011CQGra..28i4013H} in this work and { detect $N_{\text{found}} \sim 108000$ events for selection-effect calculation.}
In our analysis, we utilize software packages \texttt{gwpopulation}~\cite{2019PhRvD.100d3030T} to generate GW parameters and employ \texttt{Bilby}~\cite{Ashton:2018jfp,Romero-Shaw:2020owr} to compute the Fisher matrix and estimate hyperparameters. 
Comparing the normalized distributions between all injected events and those of the found events gives the selection fraction, which depends on mass, distance, and peculiar velocity, see Eq.~(\ref{eq:selection}).
Subsequently, we use \Eq{eq:likelihood} to derive the hyper posterior for the peculiar velocity distribution.

\begin{figure}[htbp!]
	\centering
	\includegraphics[width=\textwidth]{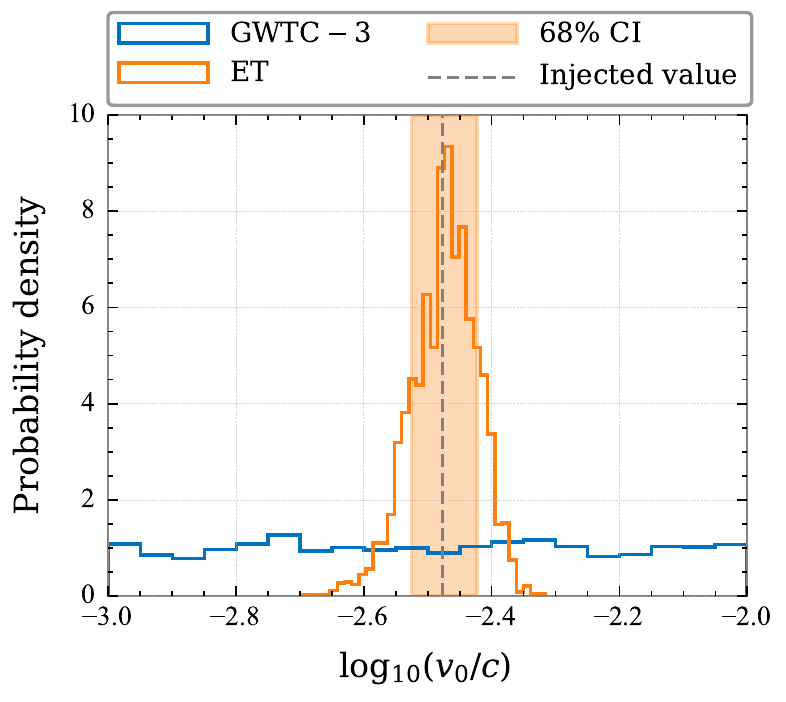}
	\caption{\label{H0_joint} The one-dimensional marginal posterior distribution for $v_0$. The grey dash line is the injected value ${\rm{log}_{10}}(v_0/c) =-2.48  $ (corresponding to $v_0=1000\,{\rm{km\,s^{-1}}}$)  for simulated events of ET. The blue solid and orange solid lines represent the posterior of $v_0$ derived from GWTC-3, and ET, respectively. The orange shaded area indicates the 68\% credible interval, namely 1$\sigma$ region.}\label{fig:gwtc_et_post_v0}
\end{figure}

Figure~\ref{fig:gwtc_et_post_v0} shows the marginal posterior distribution of $v_0$, where the orange histogram is obtained using $10^3$ golden BBH events detected with ET, and the grey dashed line indicates the injection value. Here, a uniform prior in $[10, 4000]\,{\rm{km\,s^{-1}}}$ is used for $v_0$.
The one-dimensional marginal posterior distribution of $v_0$ using $10^3$ golden BBH events detected with ET is shown as the orange histograms in Figure~\ref{fig:gwtc_et_post_v0}. 
Where the grey dash line indicates the injection value for $v_0$, 
and uniform priors for $v_0$ is $[10, 4000]\,{\rm{km\,s^{-1}}}$.
The posterior of $v_0$ is {$1014^{+119}_{-117}\,\,\rm{km\,s^{-1}}$ }, giving a relative measurement error of $\sim$ 10\%.

\section{\label{sec:discussion}Conclusion and discussion}

GW provides a new method to study the peculiar velocity distribution of BBHs. In this paper, we investigate how the detector-frame black hole mass distribution, which encodes information about cosmic expansion and peculiar velocity, can be utilized to infer the peculiar velocity distribution without direct redshift measurement. We assume a $\Lambda$CDM cosmological model for peculiar velocity distribution parameters estimation and use BBHs from GWTC-3. Our result indicates that for the current BBH events from GWTC-3, the constraint on $v_0$ is weak and uninformative, mainly due to the small BBH sample and the large uncertainty of mass and luminosity distance. However, we find that a one-year conservative observation with ET ($10^3$ golden BBH events) will measure $v_0$ with a relative uncertainty of $\sim 10\%$, assuming current knowledge of the black hole mass function, and the merger rate evolution. A better understanding of the black hole mass function and the merger rate evolution, which is plausible in the era of the third-generation gravitational-wave detectors, will achieve more precise constraints on the peculiar velocity distribution.

In this work, we provide a framework that allows simultaneous inference of peculiar velocity distribution, mass function, and merger rate evolution parameters for black hole samples without electromagnetic counterparts or galaxy catalogs.
We expect favorable constraints on the peculiar velocity distribution of the BBH events with $\mathcal{O}$(thousands) of GW sources and more accurate measurement as detector sensitivity improves by ET in the coming years. The accurate peculiar velocity distribution can aid in comprehending the formation, dynamics, evolution, and merging history of black holes. In the future, with the increased BBH sample and improved detectors, our approach might provide a robust inference for the peculiar velocity distribution of black holes.

Note that the peculiar velocity generally depends on the masses of BBHs, and this work has not considered the potential impact of kick velocity. The kick, arising from the emission of gravitational radiation during the merging of BBHs, leads to a recoil in the remnant black holes. This process is a complex phenomenon that necessitates the utilization of numerical relativity (NR) simulations, which provide essential insights into the evolution of BBH mergers and furnish significant results regarding the post-merger recoil or kick velocity of the remnant black holes~\cite{ 2016PhRvL.117a1101G, 2019PhRvR...1c3015V, CalderonBustillo:2018zuq}. Recent advancements in the field have introduced NR surrogate waveform models like {\tt{NRSur7dq4}}~\cite{2019PhRvR...1c3015V}, offering a means to constrain the kick velocity of the final black holes, see e.g.~\cite{2021ApJ...914L..18D,2022PhRvL.128s1102V,2022PhRvL.128c1101V}. 
However, it is crucial to clarify that our current study adopts a different approach. We rely on combined posterior samples derived from IMRPhenom~\cite{Thompson:2020nei,Pratten:2020ceb} and SEOBNR~\cite{Ossokine:2020kjp,Matas:2020wab} waveforms. Unlike NR simulations, these waveforms do not inherently include information about the recoil experienced by merged black holes. This limitation restricts our ability to infer the kick velocity as a function of binary masses. 
Consequently, we have chosen to focus solely on the inspiral phase and neglect the potential influence of the kick on the peculiar velocity.

Furthermore, peculiar velocities observed in GW events may arise from additional astrophysical effects beyond the linear momentum dissipated by GW emission. Typically, these alternative sources tend to outweigh the velocity imparted to the center of mass of the binary due to GW-induced linear momentum dissipation.
For instance, Ref.~\cite{Blanchet:2005rj} indicates a net recoil of approximately 22 ${\rm km\,s^{-1}}$ experienced by the binary during its inspiral phase toward the innermost stable circular orbit. However, as highlighted in Ref.~\cite{LIGOScientific:2017adf, Hjorth:2017yza}, the host galaxy of GW170817, NGC 4993, exhibits a peculiar motion estimated at $\sim$ 300 ${\rm km\,s^{-1}}$. The velocity of the host galaxy surpasses the contribution from the dissipation of linear momentum by GWs. 
Consequently, during the early stages of the inspiral, specifically at a reference frequency of 20 Hz, the velocity contributed by GW emission can be safely neglected.
This justifies the validity of the current approach, assuming that the peculiar velocity and mass distributions are separable.

\appendix
\label{sec:append}
	
\acknowledgments
We thank the referee for useful suggestions and comments, and also Xing-Jiang Zhu, Zong-Hong Zhu, and Shen-shi Du for their valuable discussions.
ZQY is supported by the National Natural Science Foundation of China under Grant No. 12305059 and the China Postdoctoral Science Foundation Fellowship No. 2022M720482.
ZCC is supported by the National Natural Science Foundation of China (Grant No.~12247176 and No.~12247112) and the China Postdoctoral Science Foundation Fellowship No.~2022M710429. 
LL is supported by the National Natural Science Foundation of China (Grant No.~12247112 and No.~12247176) and the China Postdoctoral Science Foundation Fellowship No. 2023M730300. 
ZY is supported by the National Natural Science
Foundation of China under Grant No. 12205015 and the supporting fund for young researcher
of Beijing Normal University under Grant No. 28719/310432102.
This research has made use of data, software and/or web tools obtained from the Gravitational Wave Open Science Center~\cite{web_gw_opensince}, a service of LIGO Laboratory, the LIGO Scientific Collaboration, the Virgo Collaboration, and the KAGRA Collaboration. 
	
\bibliographystyle{JHEP}
\bibliography{ref}

\newpage
\appendix
\section{\label{append} Posterior distributions of all hyperparameters}

This appendix shows the posteriors of all the peculiar velocity
distribution and population parameters using BBH events in GWTC-3. The corner plots are generated utilizing the \texttt{corner}~\cite{corner} package.
\begin{figure}[htbp!]
	\centering
	\includegraphics[width=1\textwidth]{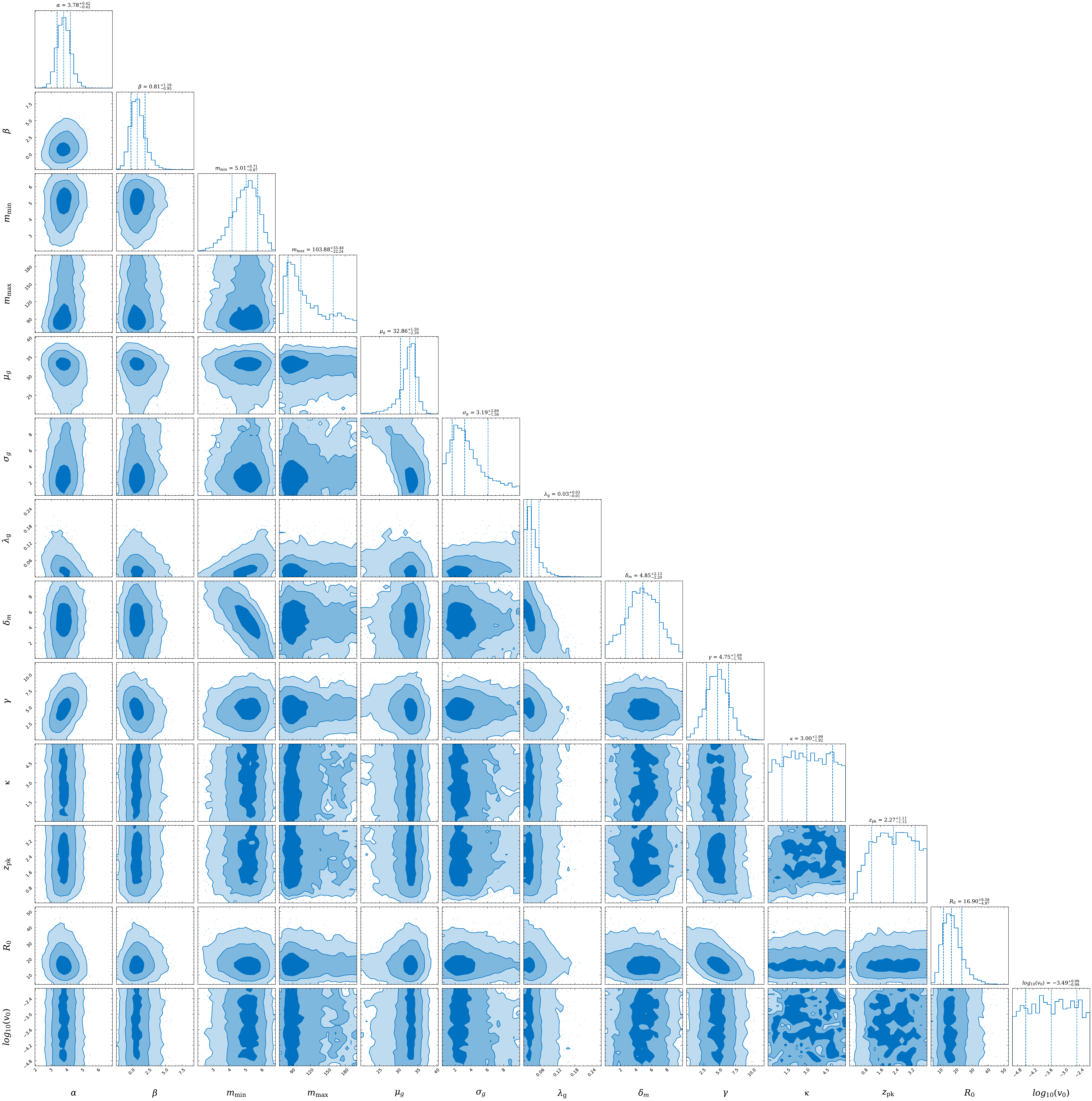}
	\caption{\label{fig:all_para} Posterior distributions of all the hyperparameters, including {\tt{POWER LAW + PEAK}} mass function parameters ($\alpha,\beta,m_{\rm{min}}, m_{\rm{max}},\mu_g, \sigma_g, \lambda_g, \delta_m$), merger rate evolution parameters ($\gamma, \kappa, z_{\rm{pk}}, R_0$), and the peculiar velocity distribution parameter $v_0$.}
\end{figure}
\end{document}